\documentclass[preprint,12pt]{elsarticle}
\usepackage{float}
\usepackage{graphicx}
\usepackage[latin1]{inputenc}
\usepackage{amssymb}

\newcommand{\be}{\begin{equation}}
\newcommand{\ee}{\end{equation}}
\newcommand{\bea}{\begin{eqnarray}}
\newcommand{\eea}{\end{eqnarray}}
\newcommand{\ba}{\begin{array}}
\newcommand{\ea}{\end{array}}

\newcommand{\bit}{\begin{itemize}}
\newcommand{\eit}{\end{itemize}}
\newcommand{\ben}{\begin{enumerate}}
\newcommand{\een}{\end{enumerate}}

\begin{document}


\begin{frontmatter}

\title{Grand Minima Under the Light of a Low Order Dynamo Model}

\author[a]{D\'ario Passos}
\ead{dariopassos@ist.utl.pt}
\author[a,b]{Il\' \i dio Lopes}
\ead{ilidio.lopes@ist.utl.pt}

\address[a]{CENTRA, Departmento de  F\'\i sica, Instituto Superior T\'ecnico, Av.
Rovisco Pais, 1049-001 Lisboa, Portugal}
\address[b]{Departamento de F\'\i sica, Universidade de \'Evora, Col\'egio Ant\'onio Luis
Verney, 7002-554 \'Evora - Portugal}

\begin{abstract}

In this work we use a low order dynamo model and study under which conditions can it
reproduce solar grand minima. We begin by building the phase space of a proxy for the
toroidal component of the solar magnetic field and we develop a model, derived from mean
field dynamo theory, that gives the time evolution of the toroidal field. This model
is characterized by a non-linear oscillator whose coefficients retain most of the physics
behind dynamo theory. In the derivation of the model we also include stochastic oscillations
in the $\alpha$ effect. We found evidences that stochastic fluctuations in $\alpha$ effect
can trigger grand minima episodes in this model under some considerations.
We also explore other ways of creating grand minima by looking into the physical mechanisms
that compose the coefficients of the oscillator. The balance between meridional circulation
and magnetic diffusivity as well as the field intensification by buoyancy driven instabilities,
might have a crucial role in inducing grand minima.

\end{abstract}


\end{frontmatter}


\section{Introduction}

The Sun presents variability in several time scales, ranging from days to decades. The
mechanisms behind this variability are still poorly understood although the common ground for
most of them involve magnetic fields and turbulence.
One of the main signatures of the solar magnetic activity is the cyclic formation of spots
in the solar photosphere, usually known as sunspots. This sunspot cycle is also accompanied by
changes in the solar spectrum. However, this cyclic activity is not regular since the peak
amplitude and duration of the cycles changes with time.
Sometimes these cycles even appear to be completely suppressed during long periods of time,
giving rise to a specific kind of solar phenomena, the so called grand minima. In these periods
the Sun appears to be in a very calm state, almost not exhibiting any sign of magnetic activity
(spots, flares, etc...). The origin of these long periods of "solar inactivity" is still
unknown and pose interesting scientific challenges.

It is believed that the solar magnetic cycle has its origin in a dynamo process that operates
in the convection zone and converts kinetic energy from the solar plasma flows into magnetic
energy. When we have a grand minimum, the dynamo changes its operation regime and apparently
shuts off for some time. The most famous grand minima that is registered is the Maunder Minima
which occurred between the years of 1645 and 1715 (\citet{Eddy1976}). During this period, although
there were no apparent signs of activity, several studies indicate that the dynamo was still
operating (e.g. \citet{Beer1998}, \citet{Miyahara2004}).

To fully understand the intrinsic physics behind the dynamo one needs to resort to the
magneto-hydrodynamic theory (MHD) which can be a very complex and difficult subject to fully
grasp (\citet{Charbonneau2005}). Thankfully nowadays the fast development of computer science
allows us to study these complex equations through the implementation of numerical
dynamos. These "tools", presently represent the best way of studying the processes involved
in the dynamo operation. Some encouraging results on possible mechanisms behind grand minima
have  been presented in the last years (\citet{Charbonneau2000}, \citet{Charbonneau2004},
\citet{Moss2008}, \citet{Brandenburg2008}, \citet{Choudhuri2009}).

As an alternative to MHD some authors, mainly in the 1990's, used low-dimensional chaotic
systems to describe the behavior of the solar magnetic cycle (e.g. \citet{Ruzmaikin1981},
\citet{Ostriakov1990}, \citet{Serre2000}). Low order models are simpler to compute but
their interpretation can sometimes be tricky. Since they involve the collapse of the number
of variables into a space with lower variables number, some information might be lost during
the transformation. A low-order systems can be seen as a "projection" of a higher order
system where the final result depends on the initial system and the "projection method" used.
Due to this, it should be noted that reduced-order systems are often abstract representations
which can loose physical meaning (\citet{Antoulas2001}).
By paying attention to these sensible points, dynamical system analysis involving low order models
has proved to be a great tool in science. In more recent years, work developed by, e.g.
 \citet{Mininni2001}, \citet{Pontieri2003}, \citet{WilmotSmith2005}, \citet{PassosLopes2008},
 \citet{Lopes2009} suggests that within certain conditions, some of the observed properties
 of the solar magnetic field can be explained by low order dynamical models.

In this work we intend to give a side perspective to the possible grand minima origins using
a low-order dynamical system derived from dynamo theory. Although more limited than computational
models this approach might be useful to build up intuition on physical processes.

The model we use here is analogous to the one presented in \citet{PassosLopes2008} and describes
the evolution of the toroidal component of the solar magnetic field. Looking
to the model's parameters we intend to study under which conditions can it reproduce grand
minima. In order to compare this model with observational results, we use the sunspot number
to build a proxy for the toroidal component and we look for the effects of grand minima
in the phase space of this proxy. This gives us an experimental signature for grand minima that
we should be able to reproduce with our model. We finish this present work with a discussion about
the results obtained.

\section{Data and Grand Minima}

In order to study grand minima, we need to use solar activity records that go back in time
to at least 1610, in order to include one of the most relevant grand minimum, the Maunder Minimum.
For that purpose, we use the revised Sunspot Group Numbers (monthly averages), $Rg$, from
\citet{Hoyt1997} and available at NOAA database
\footnote{http://www.ngdc.noaa.gov/stp/SOLAR/ftpsunspotnumber.html}. After 1995 the time series
is completed with the International Sunspot Number.

\begin{figure}[htb!]
    \centering
    \includegraphics[scale=0.5]{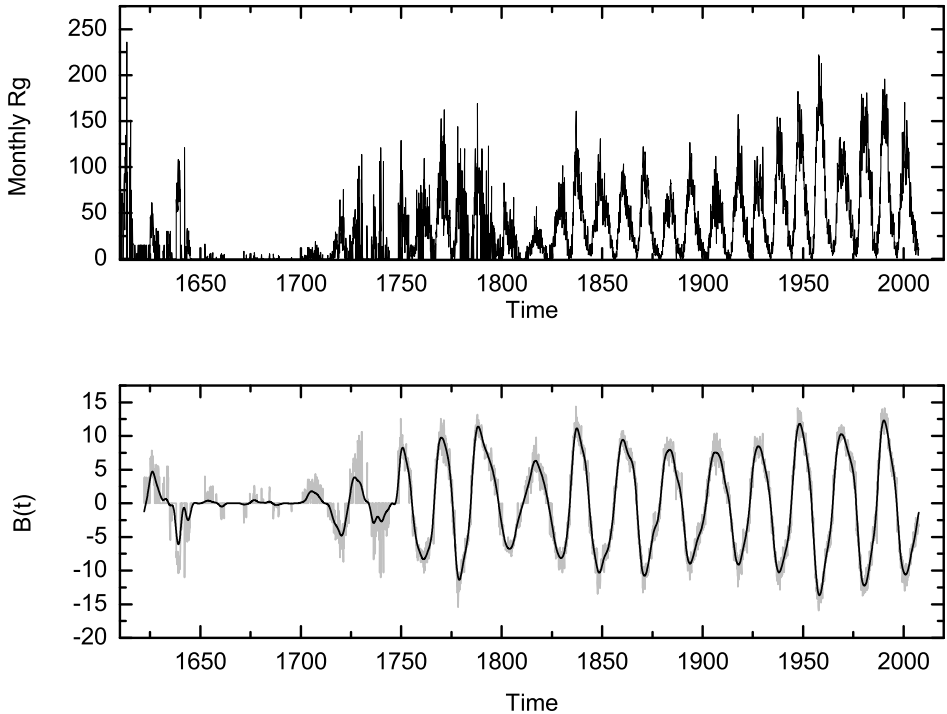}
    \caption{Top: Group Sunspot Number. Bottom: In black we have the built proxy for the
    toroidal field, $B(t)$, superimposed to $\pm\sqrt{Rg}$ in gray.}
    \label{fig-1}
\end{figure}

As it is generally accepted, sunspots are a consequence of the toroidal magnetic field
inside the convection zone, more specifically we can say that the sunspot number is proportional
to the magnetic energy ($\propto B^2$) beneath the photosphere.
Thus, we use $Rg$ to build a proxy for this component of the field simply by assuming that
$B(t)\propto\pm\sqrt{Rg}$. To account for field reversals we change the sign of $B(t)$ by hand for
every sunspot cycle. To identify solar minima we used a low pass filter and selected the lowest
values of the data series. Since identifying individual cycles in the period of the Maunder Minimum
is very difficult, we decided to divide it into four separate "suppressed"\, cycles. At this point we
would like to note that since the amplitude of $Rg$ during this period is very small, for the purpose
of this work, a different choice would not have made an impact.
In order to get the average behavior of the time series and eliminate "fast"\, transients (lower than
2.6 years), the proxy data is smoothed using a FFT filter (see figure (\ref{fig-1})).
At this point we would like to note that the use of sunspots to build the $B(t)$ proxy and the methodology
applied, is going bind us to a characteristic dynamo scale whose behavior can, in principle, be
reproduced by a low order model.

As observed by \citet{Polygiannakis1996} a phase space reconstruction of the sunspot number
hints that its behavior might be described by a non-linear oscillator. We pursue this idea
but instead we use the proxy that we built. In order to construct our phase space, the numerical
derivative, $\mathrm{d}B/\mathrm{d}t$, is computed using a time step of twelve months.

\begin{figure}[htb!]
    \centering
    \includegraphics[scale=0.6]{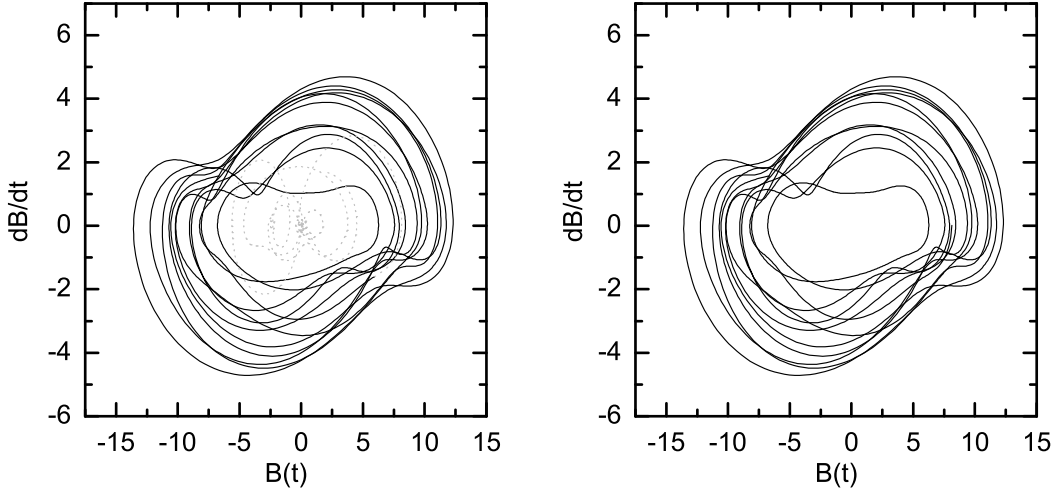}
    \caption{In this figure it's presented the constructed phase space of $B(t)$ without the
    Maunder Minimum (right), and with it (left) in light gray dots to clearly distinguish it.}
    \label{fig-2}
\end{figure}

Despite a small randomness, the trajectories of $B(t)$ in the phase space appear to be stable,
and seem to indicate that the solution for this oscillator is some kind of attractor. The only
moment that the oscillator seems to seriously deviate from its "natural" action area (it collapses)
is during the Maunder Minimum period (from approx. 1650 to 1720), depicted in gray in figure
(\ref{fig-2}). This is the experimental signature of grand minima that we will try to reproduce
with the low-order model.

\section{Low order dynamo model with a stochastic $\alpha$ effect}

In order to find an expression for a possible non-linear oscillator that might explain the
behavior presented in the phase space of the toroidal field depicted in figure (\ref{fig-2}),
we follow the ideas of \citet{Mininni2001} and \citet{Pontieri2003}. The model presented
here is also discussed in \citet{PassosLopes2008} although with a different objective and
derivation. Instead of proposing a purely mathematical inspired \textit{ad hoc} expression
for the oscillator, we intend to derive it from dynamo equations. This will allows us to
connect physical mechanisms from the dynamo with coefficients in the oscillator's expression.

We start by writing the equations for a mean field axisymmetric dynamo as shown in
\citet{Charbonneau2005}. These equations give us the evolution of the mean solar magnetic
field, $\bar{\textbf{B}}$, classically decomposed into its toroidal and poloidal components,
$\bar{\textbf{B}}=\textbf{B}_\phi + \textbf{B}_p$ with
 $\textbf{B}_p = \nabla \times (A_p \hat{e}_\phi$).

\bea
    \frac{\partial B_\phi}{\partial t}&=& \eta \left( \nabla^2 -
    \frac{1}{\bar{r}^2} \right) B_\phi + \frac{1}{\bar{r}} \frac{\partial
    (\bar{r}B_\phi)}{\partial r} \frac{\partial \eta}{\partial r} - \bar{r}
    \textbf{v}_p \cdot \nabla \left(\frac{B_\phi}{\bar{r}}\right) -
    B_\phi \nabla \cdot \textbf{v}_p \nonumber \\
    && + \bar{r}\left[\nabla \times (A_p \hat{e}_\phi) \right] \cdot \nabla \Omega
    \label{eq-1} \\
    \frac{\partial A_p}{\partial t} &=& \eta \left( \nabla^2 - \frac{1}{\bar{r}}
    \right) A_p - \frac{\textbf{v}_p}{\bar{r}} \cdot \nabla (\bar{r} A_p),
    \label{eq-2}
\eea
where we have $\bar{r}=r \sin \theta$, $\nabla \Omega$ represents the differential rotation of
the Sun, $\textbf{v}_p$ is the flow in the meridional plane and $\eta$ is the magnetic diffusion.
For simplification we will assume that $\eta$ is a constant in all of the convection zone
( $\partial \eta / \partial r = 0$) and that the plasma is incompressible. We then get

\bea
    \frac{\partial B_\phi}{\partial t} &=& - \bar{r} \, \textbf{v}_p \cdot \nabla
    \left(\frac{B_\phi} {\bar{r}}\right) + \bar{r}\left[\nabla \times (A_p \hat{e}_\phi)
    \right] \cdot \nabla \Omega\ + \eta \left(\nabla^2 - \frac{1}{\bar{r}^2}\right) B_\phi
    - \Gamma(B_\phi) B_\phi\,\,\, ,
    \label{A-eq-1} \\
    \frac{\partial A_p}{\partial t} &=& -\frac{1}{\bar{r}} \,\textbf{v}_p \cdot
    \nabla \left(\bar{r}\,A_p\right) + (\alpha_0 + \alpha_r(t)) B_\phi + \eta \left( \nabla^2
    - \frac{1}{\bar{r}^2} \right) A_p \,\,\, ,
    \label{A-eq-2}
\eea
where we introduced a simple linear $\alpha$ effect in the form of $\alpha=\alpha_0 + \alpha_r(t)$
defined as having a constant part, $\alpha_0$, and a stochastic part, $\alpha_r(t)$,
that changes through time. The effect of a stochastic excitation in the $\alpha$ effect has been
studied in numerical dynamo simulations and in theoretical dynamos by several authors
(e.g. \citet{Charbonneau2000}, \citet{Brandenburg2008}, \citet{Moss2008}) and is justified by
the angle dispersion around the mean tilt presented by bipolar active regions as they emerge.
All these works present evidence that this stochastic effect might be behind grand minima phenomena,
hence, we decided to introduce it. Also, in order to account for the removal of the toroidal
field from the bottom of the convection zone by magnetic buoyancy we follow the suggestions of
\citet{Pontieri2003} and add a term, $\Gamma \sim \gamma B_\phi^2 / 8 \pi \rho$, where $\gamma$ is
a constant related to the buoyancy regime and $\rho$  is the plasma density. This
term can also work as an extra source for poloidal field due to buoyancy driven instabilities
in the apex of rising flux tube (\citet{Rempel2001}). As noted before, our sunspot derived proxy,
gives us the magnetic field average behavior for a certain scale. In order to capture phenomena
just on that scale, we truncate the dynamo equations by substituting $\nabla \rightarrow 1/l_0$,
where $l_0$ is a specific length of interaction for the magnetic fields.

After grouping terms in $B_\phi$ and $A_p$ we get

\bea
    \frac{\partial B_\phi}{\partial t} &=& c_{1} B_\phi + c_{2} A_p - c_{3} B_\phi^3
    \label{A-eq-11} \\
    \frac{\partial A_p}{\partial t} &=& c_{1} A_p + \alpha_0 \, B_\phi
    +\alpha_r(t)\, B_\phi,
    \label{A-eq-12}
\eea
where we have defined the coefficients, $c_n$, as
\bea
    c_{1}&=&\eta \left(\frac{1}{l_0^2} -
    \frac{1}{\bar{r}^2} \right) - \frac{v_p}{l_0}
    \label{A-eq-13} \\
    c_{2}&=& \frac{\bar{r} \Omega}{l_0^2}
    \label{A-eq-14} \\
    c_{3} &=& \frac{\gamma}{8 \pi \rho}
    \label{A-eq-15}
\eea

We now concentrate in creating an expression for the time evolution of $B_\phi$ since it is
the quantity represented by our proxy $B(t)$. To do so, we derive expression (\ref{A-eq-11})
in order to the time, and substitute (\ref{A-eq-12}) in it to take away the $A_p$ dependence.
After some mathematical manipulation it is possible to show that

\be
    \frac{\partial^2 B_\phi }{\partial t^2} + (\omega^2  - c_2 \alpha_r(t)) B_\phi +
    \mu (3 \xi B_\phi^2 - 1)\frac{\partial B_\phi}{\partial t} - \lambda B_\phi^3 = 0,
    \label{A-eq-23}
\ee
where $\omega^2=c_1^2 - c_2 \alpha_0$, $\mu = 2c_1$,
$\xi=c_3 / 2 c_1$ and $\lambda=c_1 c_3$ are coefficients that contain the solar physical structure
(rotation, flows, diffusivity, etc.).

\begin{figure}[htb!]
    \centering
    \includegraphics[scale=0.9]{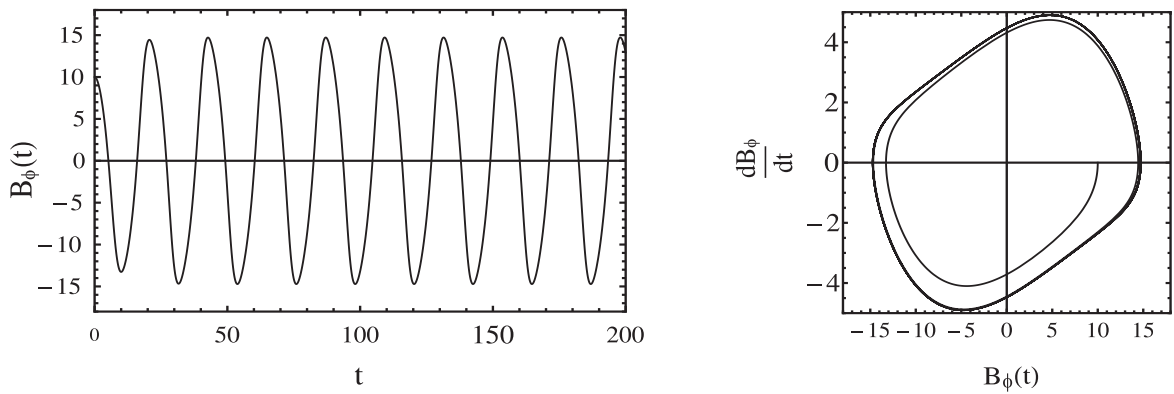}
    \caption{Solution for $B_\phi$ corresponding to the solar cycle. The evolution of the magnetic
    field (left) and the corresponding phase space (right). Solution obtained using $c_1=0.08$,
    $c_2=-0.09$, $c_3=0.001$, $\alpha_0=1$ and $\alpha_r=0$.}
    \label{fig-3}
\end{figure}

The fact that the solar magnetic field presents a cyclic behavior in time, hints that the solution
of this non-linear oscillator (van der Pol - Duffing type), in the phase space, would approximately
correspond to a closed cycle (closed trajectory) or attractor whose shape depends on the structure
parameters $\omega$, $\mu$, $\xi$ and $\lambda$. From the dynamical system's point of view each one
of these parameters will control the system in different ways.
Usually, in the classical case of this oscillator ($\alpha_r(t)=0$), $\omega$ controls the frequency
of the oscillations and the other term that also has an effect in the frequency is $\lambda$.
In the presence of $\alpha$ fluctuations, i.e. $\alpha_r(t)\neq0$, the term $c_2 \alpha_r(t)$ will
appear as a perturbation to the frequency. As for the remaining coefficients, $\mu$ controls the
asymmetry between the rising and falling parts of the cycle and $\xi$ affects directly the amplitude.
As opposite to the classic van der Pol-Duffing, in our derived oscillator these coefficients are not
independent since some of them depend on the same physical quantities. This interdependency will
eventually constrain the solution's space. In figure (\ref{fig-3}) it is presented a solution that
corresponds to the solar cycle in which the magnetic field oscillates changing sign every 11
years approximately.

In a first approximation, we can assume that the physical quantities involved in the structure parameters
change in a time scale much longer than the magnetic field itself. Thus, if we look into what might change
in these parameters that can create grand minima (a collapse to the center in terms of the phase space),
one might derive some clues about the intervening physical mechanisms.

\subsection{Obtaining Minima with $\alpha_r$?}

By analyzing equation (\ref{A-eq-23}) we can see that the term $(\omega^2 - c_2 \alpha_r(t))$
is going to introduce variations in the system due to the random fluctuations of $\alpha_r(t)$. For
simplification lets call this term $\Phi_r(t)$ and study its impact in the solution.
%
%

\bea
    \Phi_r(t) = \omega^2  - c_2 \alpha_r(t) &=& c_1^2 - c_2 (\alpha_0 +  \alpha_r(t)),
    \label{A-eq-24}
\eea

\begin{figure}[htb!]
    \centering
    \includegraphics[scale=0.55]{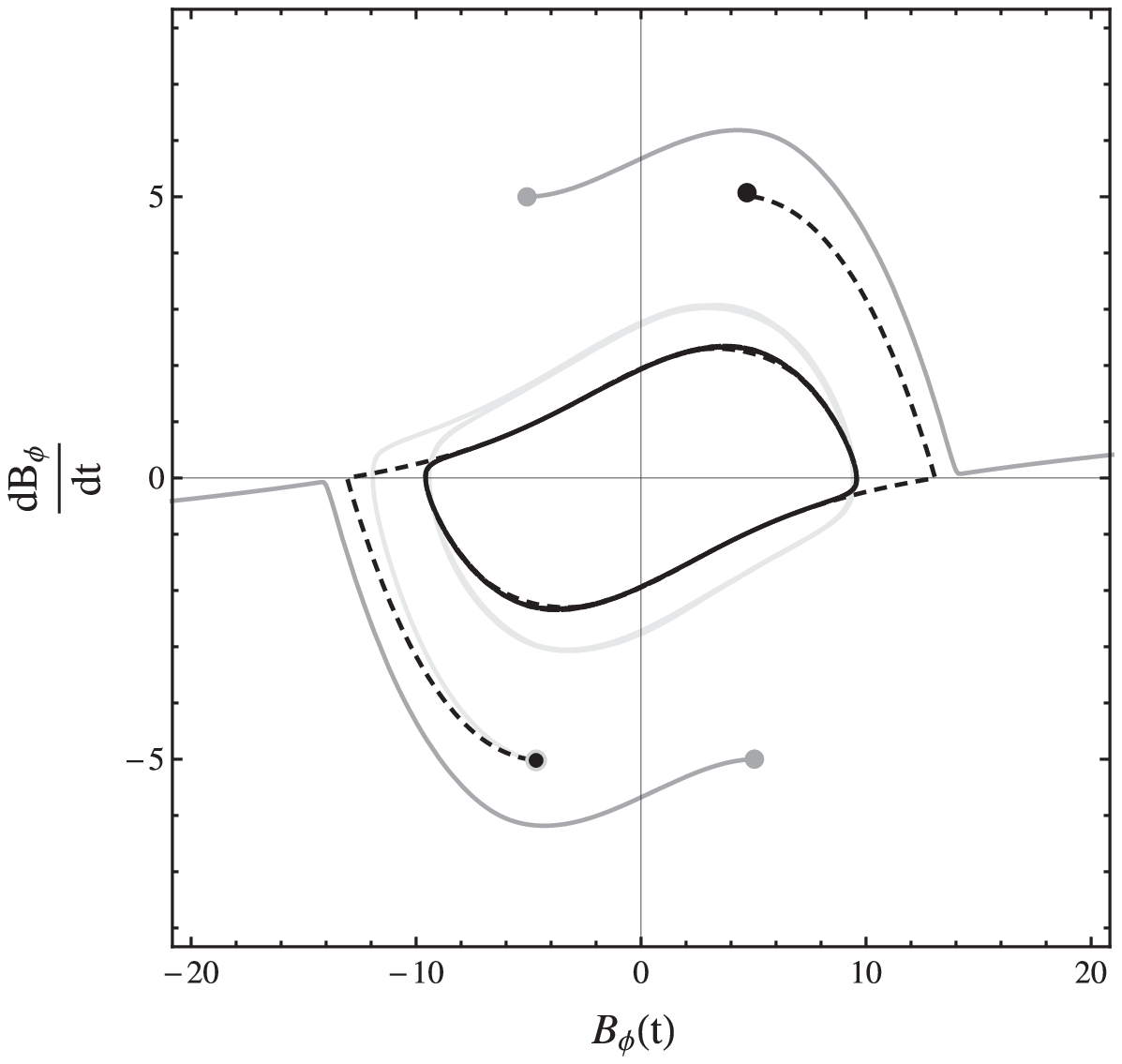}
    \includegraphics[scale=0.8]{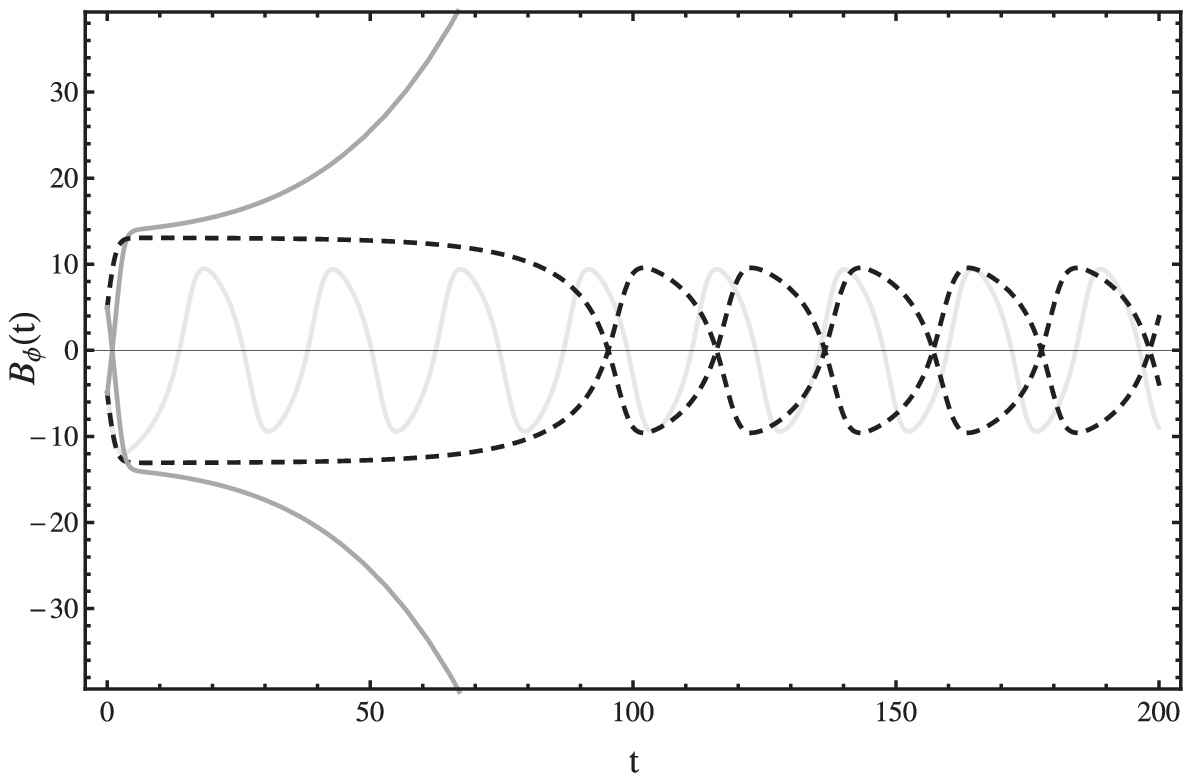}
    \caption{Representation of the phase space (left) of equation (10) with $c_1$ = 0.0933,
    $c_2$ = -0.0372, $c_3$ = 0.00287, $\alpha_0=1$ and $\alpha_r=0$ for four different initial conditions (dots).
    The light gray trajectory represents a stable solution with $c_2 = -0.09$. The time evolution
    of the $B(t)$ for the same conditions is presented on the right. }
    \label{fig-4}
\end{figure}

If one allows random $\alpha_r(t)$ fluctuations to have values between $\pm \alpha_0$ then the
solution's space for $\Phi_r(t)$ will contain an interval of values where the system is stable
and another where the system is unstable. In the later  interval the regular oscillations
disappear and the solutions grow positive or negative depending on the value of the other
coefficients and the point where the system is at that moment.
Figure (\ref{fig-4}) shows a pictorial example of this.

In this figure, for the two black dashed trajectories, the system finds a stable solution after
spending some time in an "almost saturated state". After the system reaches the solution given by
 the attractor it becomes completely stable. This happens for some super critical values
of $\Phi_r(t)$. On the other hand, for the same values but starting from a different point in
the phase space (different initial conditions), the dark gray solid trajectories become
unstable, drifting away from the attractor and growing indefinitely.

\begin{figure}[htb!]
    \centering
    \includegraphics[scale=1]{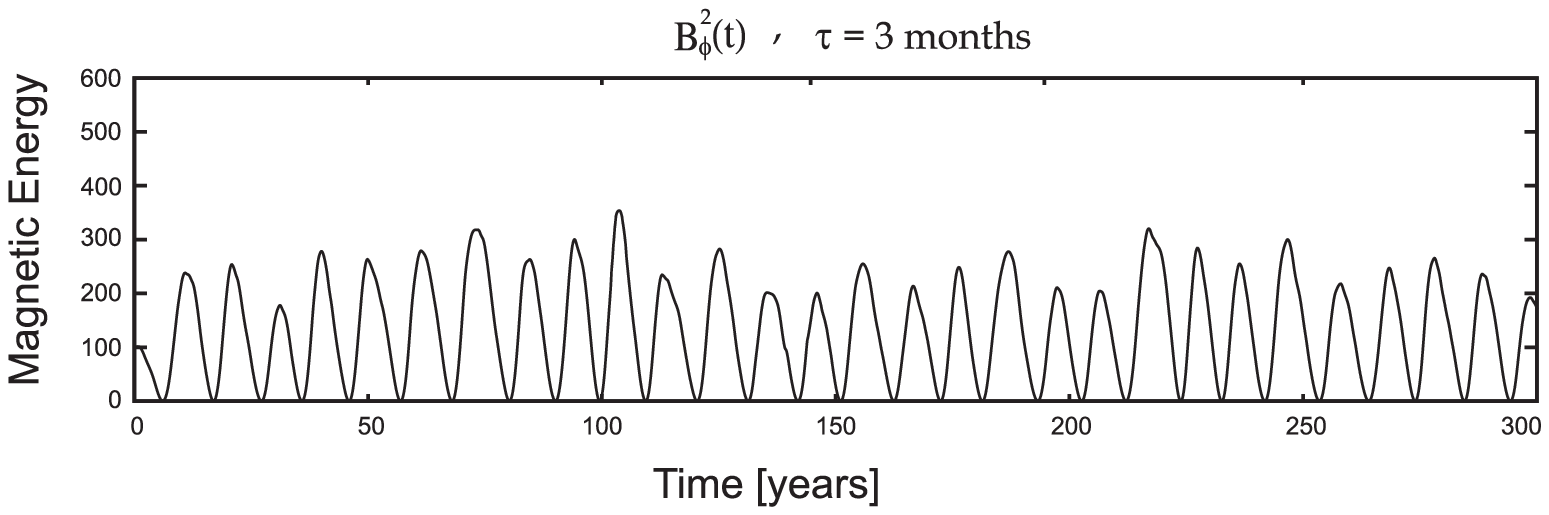} \\
    \includegraphics[scale=1]{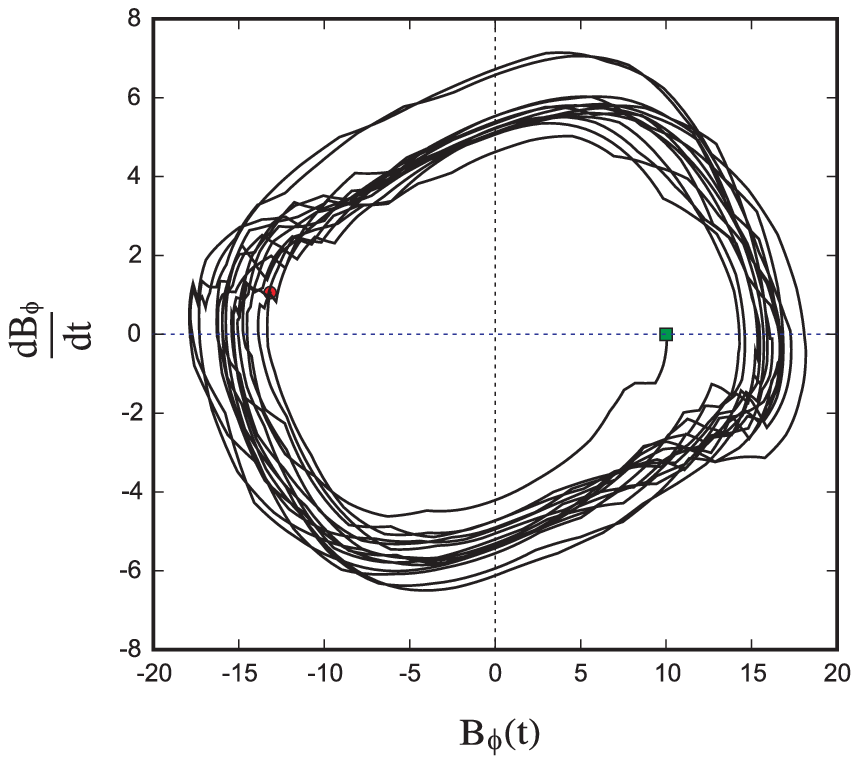}
    \caption{Representation of $B_\phi^2$ (top) and corresponding phase space (bottom). In this simulation
    we used $c_1=0.08$, $c_2=-0.09$, $c_3=0.001$, $\alpha_0 + \alpha_r \in \{0,2\}$. The green square
    represents the beginning of the simulations and the red dot, the end.
    }
    \label{fig-5}
\end{figure}

In order to simulate the effect of the fluctuations, we have to introduce a correlation
time, $\tau$. This time between fluctuations depends on the type of $\alpha$ effect
chosen (e.g. Babcock-Leighton's or Parker's) and its position (e.g. photosphere, bulk of the
convection zone, etc.). Buffeting due to turbulence during the rise of flux tubes or the decay
of bipolar active regions give us an interval between 0,5 and 4 months for the fluctuations.
In figure (\ref{fig-5}) we present the solution obtained from our model to a variation of
200\% in the $\alpha$ effect with $\tau=3 $ months.

The result shows that the fluctuations can reduce the amplitude of some cycles. In simulations over
longer periods of time, it is possible to find several weak consecutive cycles that we can interpret
as a grand minimum. This result is compatible with the one presented by \citet{Charbonneau2000},
obtained through the use of a more sophisticate computational dynamo model.
From flux tubes simulations we know that if $B_\phi$ stays bellow a certain threshold, then it will
not be stable enough to survive the rise through the convection zone and it will not produce sunspots.
If the weak cycles present in the simulation are bellow this threshold then they might be responsible
for grand minima periods. One should also note that if the correlation time and/or the level of
fluctuations increases too much then the probability for the system to be in an unstable trajectory
for a long period will also increase, originating solutions that are non solar like.

The phase space of the simulated $B_\phi$ resembles that of our built proxy $B(t)$ exhibiting almost
the same level of variability. Nevertheless, among all simulations performed, no collapse in the phase
space was found.

At this point the only conclusion that we can derive is that, in this model, a linear $\alpha$ effect
with random fluctuations might be responsible for grand minima, but it does not reproduce the collapse
of the phase space found in our proxy $B(t)$.

\subsection{Obtaining Minima by maximizing $\xi$}

A way of making the trajectories in the phase space to collapse, is to increase the $\xi$ parameter.
Until a certain threshold, the higher $\xi$ is, the lower is the amplitude of $B_\phi$. At this point,
it is useful to remember that the construction of $B(t)$ involves an unknown proportionality with $Rg$.
This means that although in the phase space we built we can observe a full collapse of the trajectories,
in order not to produce sunspots the field doesn't need to go all the way to zero as was noted before.
So, in principle, we could look into the physical processes contained in $\xi=\frac{c_3}{2 c_1}$
and check which one can maximize this coefficient.
\bea
    \xi &=& \frac{\frac{\gamma}{8 \pi \rho}}{ 2 \eta \left(\frac{1}{l_0^2} -\frac{1}{\bar{r}^2}\right)
    -\frac{2 v_p}{l_0}}\,\,\,.
    \label{A-eq-24}
\eea

\begin{figure}[htb!]
    \centering
    \includegraphics[scale=1]{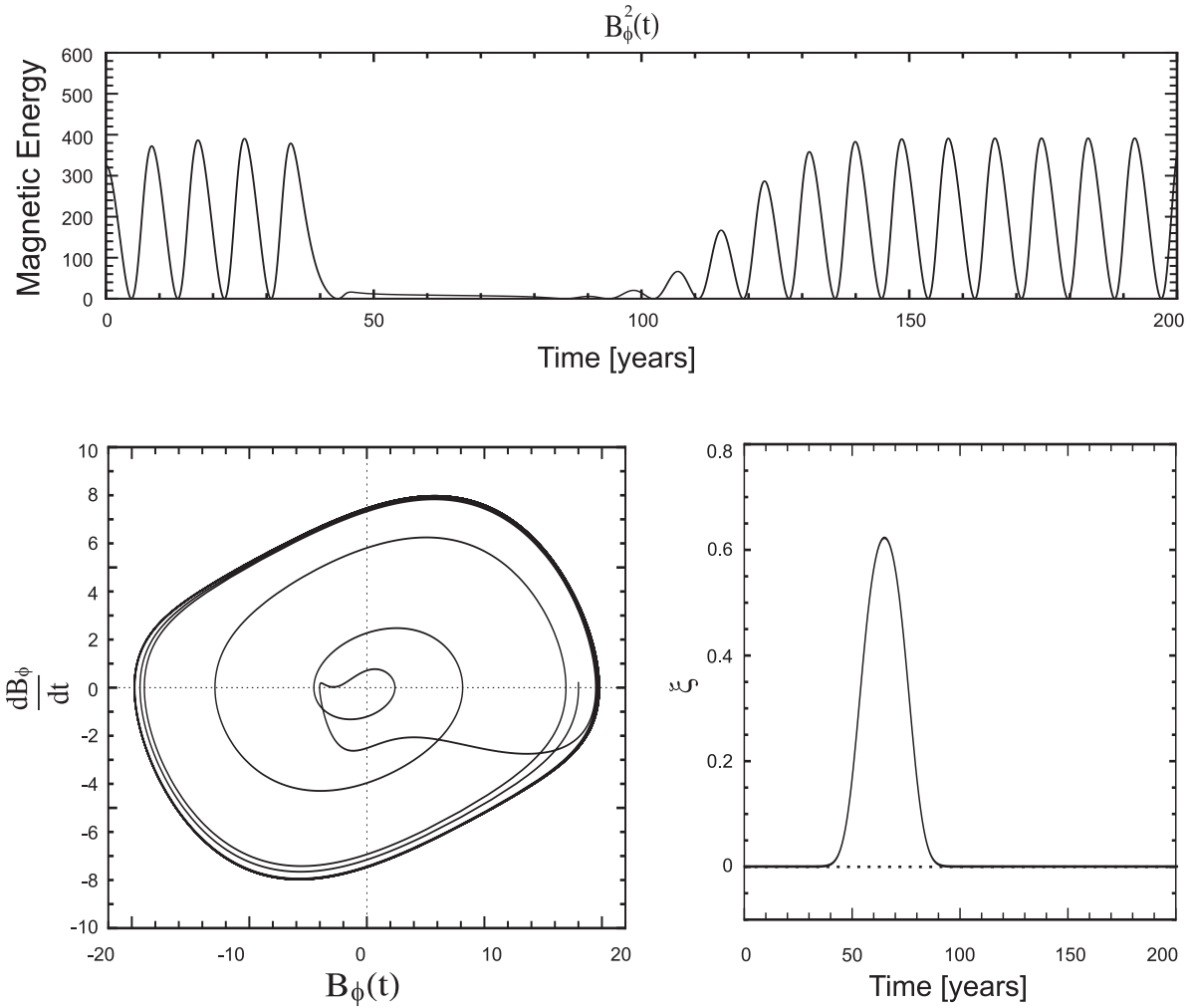}
    \caption{The evolution of $B_\phi^2$ (top) and correspondent phase space (bottom left) for a $\xi$
    profile that varies between 0.005 and 0.62 (bottom right).
    }
    \label{fig-6}
\end{figure}

In this scenario three physical mechanisms come to play in the amplitude of the field: intensification
of the field due to buoyancy instabilities ($\sim \gamma$), magnetic diffusivity and meridional circulation
amplitude. Several interconnections between these mechanisms can be arranged in order to
increase $\xi$. Since these mechanisms can also be affected by stochastic or other forcing factors
(e.g. the meridional circulation can be affected by the field feedback, or $\eta$ can incorporate a quench)
their inter-relations can create the necessary conditions to increase $\xi$ and create a grand minimum.
For example, if we consider that $c_3$ is constant, then to create an increase/decrease in the
amplitude of $B_\phi$ is a change in the balance between magnetic diffusion and the meridional flow.
This may happen if the dynamo shifts from an advective to a diffusive regime or vice-versa.
Figure (\ref{fig-6}) shows the response of the system to a temporal increase of $\xi$.
This increase makes $B(t)$ collapse in the phase space, creating a signature analogous to the one
depicted in our proxy data.

This result is also compatible with the work of \citet{Usoskin2000} in the sense that
the minimum is triggered almost in just one cycle but it comes out of it gradually.
Using a numerical dynamo code, \citet{Choudhuri2009} obtained a similar result after
inducing a minimum by manually lowering the poloidal field at the surface at the end
of one cycle.

\section{Discussion}

Grand minima are episodes of solar "inactivity"\, that remain basically unexplained.
Some clues about the physical mechanisms that might trigger these episodes of calmness in
the Sun have been found through the use of computational solar dynamos. In this work we
intend to present a different perspective on this subject by exploring a simple
analytical model previously created to explain characteristics in the solar cycle.
By analyzing the sunspot number time series, some information about
the latest grand minima can be recovered. With this in mind, we use this time series to
create an experimental proxy for the toroidal component of the solar magnetic field, $B(t)$,
construct its phase space and look for the signature of grand minima in this phase space.
The second step is to, based on dynamo theory, develop a simple model aimed to explain
the average evolution to the toroidal field component. We modify the low-order model found
in  \citet{PassosLopes2008} by adding stochastic fluctuations to the $\alpha$ effect and we
explore under which conditions can it reproduce physical solutions that resemble the
grand minima signature found in the experimental data.
In the absence of fluctuations in the $\alpha$ effect the model easily reproduces
a solution analogous to the solar cycle for a suitable choice of parameters $c_n$ (see
figure (\ref{fig-1})). The steeper rising and moderate decrease in the amplitude that is
found in the sunspot cycle is an intrinsic characteristic of the model itself.

As seen in section 3.1, in this simplified model, the presence of a stochastic
component in a linear $\alpha$ effect does not produce a signature comparable to
grand minima in the phase space. Nevertheless, this does not mean that the stochastic
fluctuations cannot induce grand minima. If the amplitude of the weak cycles found in
our simulations is bellow the threshold needed to create sunspots then the $\alpha$
fluctuations are effectively triggering minima.
Although this result is in apparent agreement with solutions found using more sophisticated
computational dynamo models, the direct comparison is not trivial. In most of the
computational  models a non-linear $\alpha$ effect that incorporate quenching terms and/or
different spacial locations is used. These non linearities introduce different physical
dependencies in the structural coefficients, $c_n$, and ultimately a different final behavior
of the system. We are currently working into incorporating such non-linear effects into our model
as well as other non-linearities in other coefficients. There is also the role of the buoyancy
term, $\gamma$, that can act as source and whose interpretation is far from trivial. This term
might be preventing our simulations to generate lower amplitude cycles when the $\alpha$
source temporarily decreases.

In our quest to recreate the collapse of the phase space from our toroidal field proxy, we
explore other possibilities in the model. A "grand minima like" signature can be obtained by an
increase of $\xi$. The physical mechanisms that can contribute to the increase of $\xi$ are
meridional velocity, magnetic diffusion and intensification of the field due to buoyancy
instabilities. The results hint that a transition between advective and diffusive regimes in
the dynamo operation might trigger this $\xi$ increase, and consequently a minimum.

At this moment, trying to infer any other physical meaning from this result would be speculative.
Much care must be taken in interpreting the quantities present in the physical coefficients, $c_n$,
since the reduction of order of the system may have changed their usual meaning. We are currently
working on this as well in order to be able to study fluctuations to the velocity field.
One advantage of the proposed model is that results can be obtained almost instantly since
it's implementation in calculus software packages such as Mathematica or Matlab is easy.
It is our belief that the model can be improved in order to give more information about the
interplay of dynamo mechanisms. For the time being, the model and method presented here should
be seen as a toy model or proof of concept for an alternative way of tackling problems associated
with the dynamo.

\bigskip

We would like to thank the editor and the two anonymous referees for their constructing critics to
the first version of this paper. They made us rethink about the problem and contributed to the overall
improvement of the work. We also would like to thank the financial support of FCT
(Funda\c{c}\~{a}o para a Ci\^{e}ncia e Tecnologia, Portugal).

\end{document}